\documentclass[hidelinks,twocolumn, prl, superscriptaddress,a4paper]{revtex4}
\usepackage{graphicx}% Include figure files
\usepackage{dcolumn}% Align table columns on decimal point
\usepackage{orcidlink}
\usepackage{csquotes}
\usepackage{float}
\usepackage{amsmath}
\usepackage{bm}% bold math
\begin{document}

\preprint{APS123-QED}

\title{Spread complexity and localization in \texorpdfstring{$\mathcal{PT}$}~-symmetric systems}%Localization does not imply suppression of complexity in the  \texorpdfstring{$\mathcal{PT}$}~  broken phase}

\author{Aranya Bhattacharya\, \orcidlink{0000-0002-1882-4177}}
\email{aranya.bhattacharya@th.if.uj.edu.pl}
\affiliation{Institute of Physics, Jagiellonian University, Łojasiewicza 11, 30-348 Kraków, Poland}
\author{Rathindra Nath Das\,\orcidlink{0000-0002-4766-7705}}
\email{das.rathindranath@uni-wuerzburg.de}
\affiliation{Institute for Theoretical Physics and Astrophysics and W\"urzburg-Dresden Cluster of Excellence ct.qmat, Julius-Maximilians-Universit\"at W\"urzburg, 
 Am Hubland, 97074 W\"{u}rzburg, Germany}

\author{Bidyut Dey\,\orcidlink{0000-0003-1527-7064}}
\email{bidyut.dey@lnf.infn.it}
\affiliation{Dipartimento di Fisica, Università della Calabria Arcavacata di Rende, I-87036 Cosenza, Italy, \\ I.N.F.N., Gruppo Collegato di Cosenza Arcavacata di Rende, I-87036 Cosenza, Italy}

\author{Johanna Erdmenger\,\orcidlink{0000-0003-4776-4326}}
\email{erdmenger@physik.uni-wuerzburg.de}
\affiliation{Institute for Theoretical Physics and Astrophysics and W\"urzburg-Dresden Cluster of Excellence ct.qmat, Julius-Maximilians-Universit\"at W\"urzburg, 
 Am Hubland, 97074 W\"{u}rzburg, Germany}

\date{\today}

\begin{abstract}
    
We present a framework for investigating wave function spreading in $\mathcal{PT}$-symmetric quantum systems using spread complexity and spread entropy. We consider a tight-binding chain with complex on-site potentials at the boundary sites.   In the $\mathcal{PT}$-unbroken phase, the wave function is delocalized.  We find that in the $\mathcal{PT}$-broken phase, it becomes localized on one edge of the tight-binding lattice. This localization is a realization of the non-Hermitian skin effect. Localization in the $\mathcal{PT}$-broken phase is observed both in the lattice chain basis and the Krylov basis. Spread entropy, entropic complexity, and a further measure that we term the \textit{Krylov inverse participation ratio} probe the dynamics of wave function spreading and quantify the strength of localization probed in the Krylov basis. The number of Krylov basis vectors required to store the information of the state reduces with the strength of localization. Our results demonstrate how measures in Krylov space can be used to characterize the non-hermitian skin effect and its localization phase transition.

% \begin{description}
% % \item[Usage]
% % Secondary publications and information retrieval purposes.
% %\item[Structure]
% %You may use the \texttt{description} environment to structure your abstract;
% %use the optional argument of the \verb+\item+ command to give the category of each item. 
% \end{description}
\end{abstract}

\maketitle

%\tableofcontents
\paragraph*{Introduction.---}

The study of complexity \cite{Nielsen_2006, Jefferson:2017sdb, Parker:2018yvk, Chapman:2017rqy, Balasubramanian:2022tpr} in quantum systems has provided new insights into the dynamical behaviour of chaotic systems and the structure of spacetime in general \cite{Brown_2016, Susskind:2014rva}. It resulted in the development of new tools and techniques in quantum many-body systems, quantum information theory, and holographic theories. Complexity was initially defined to quantify the difficulty of reaching one target state starting from an initial reference state of a quantum system \cite{Nielsen_2006}. 

%\textcolor{blue}{On the other hand, we have to cite the paper about KRYLOV complexity (the Altman paper) much more prominently here and state its main result+motivation.}

Here we use Krylov state complexity \cite{Balasubramanian:2022tpr} to study $\mathcal{PT}$-symmetric systems. The complexity in Krylov basis was first introduced to study the operator growth in the Heisenberg picture under unitary evolution and distinguish chaotic theories from integrable ones in \cite{Parker:2018yvk}. A recent advancement to this study \cite{Balasubramanian:2022tpr} was to extend the method to Schr\"odinger evolution of quantum states and to show that the choice of Krylov basis minimizes the complexity. This measure defined in the Schr\"odinger picture is dubbed as the spread complexity \cite{Balasubramanian:2022tpr}, which quantifies the spread of an initial state under the evolution of the corresponding system Hamiltonian. The spread complexity shows a distinct characteristic behaviour in chaotic, integrable and intermediate systems, providing new insights into the dynamics of information scrambling and phase transition under time evolution \cite{Caputa:2022eye, Afrasiar:2022efk, Balasubramanian:2022dnj, Caputa:2022yju, Erdmenger:2023shk, Pal:2023yik, Nandy:2023brt,Chattopadhyay:2023fob, Gautam:2023bcm, Bhattacharjee:2022qjw, Gill:2023umm, Bento:2023bjn, Aguilar-Gutierrez:2023nyk, Craps:2023ivc, Bhattacharya:2023yec, Caputa:2023vyr, Caputa:2024vrn, Huh:2023jxt, Zhou:2024rtg, Nandy:2024htc}. These concepts were extended to the non-unitary time evolution of open quantum systems \cite{Bhattacharya:2022gbz, Bhattacharjee:2022lzy, Bhattacharya:2023zqt, Bhattacharjee:2023uwx, Bhattacharya:2024uxx}. These studies exemplified how operator growth is modified in the presence of decoherence, resulting in a prolonged decay of complexity before reaching the saturation value. A further example of non-unitarity is provided by systems subject to local projective measurements, for which it was found that the spread complexity undergoes a phase transition as a function of the measurement frequency, showcasing the quantum Zeno effect \cite{Bhattacharya:2023yec}. 

The probability distribution of a state in the Krylov basis gives rise to an entropy function known as the Krylov entropy. An alternative complexity measure is given by the exponential of this entropy. In classical mechanics, the exponential of entropy is related to the accessible phase space volume. Similarly, entropic complexity estimates the number of supports of the state on the Krylov basis vectors, reflecting the system's complexity. For unitary dynamics, both the spread complexity and the entropic complexity give equivalent estimates of complexity \cite{Balasubramanian:2022tpr}.

An integral component of Krylov space orthonormalization is provided by the Lanczos coefficients. Recently, a connection between the fluctuations of the Lanczos coefficients and  Krylov localization was proposed in \cite{Rabinovici:2021qqt}. The fluctuating Lanczos coefficients act as random hopping potentials between different sites of the Krylov chain, and the Krylov localization is then similar to Anderson localization. Wave function localization in Krylov basis then corresponds to the suppression of complexity. 

In this paper, using Krylov state complexity to address the wave function spreading dynamics in the presence of localization, we consider a $\mathcal{PT}$-symmetric tight-binding model \cite{Jin:2009zc}. $\mathcal{PT}$-symmetric systems provide a natural setup to study the emergent dynamics which arise from a change of eigenspectrum from real to imaginary.  This change in the spectrum is realized by a Hamiltonian with complex conjugate potentials at the first and last sites \cite{Jin:2009zc},
\begin{equation}
    H=-J \sum_{l=1}^{N-1}\left(a_l^\dagger a_{l+1}+\text{h.c.}\right)+i\gamma \left(a_1^\dagger a_{1}-a_N^\dagger a_{N}\right).
    \label{eq:PT_Hamiltonian}
\end{equation}
The single-particle solutions of this model exhibit two distinct phases, depending on the relative values of the coupling strength, $J$ and the strength of the on-site potential $\gamma$ of the Hamiltonian. The $\mathcal{PT}$-unbroken phase has a real eigenspectrum, while the spontaneously broken phase introduces two imaginary eigenvalues. These two phases are connected by a critical point, at which the system has a two-fold degeneracy with eigenvalue zero \cite{Jin:2009zc}. In the broken phase, one pair of purely imaginary eigenvalues result in localization of the wave function on the boundary of the tight-binding lattice, which is the known \textit{non-hermitian skin effect} \cite{skin1, skinedge, skintop, Kawabata:2022biv, skinpt,Yao_2018, Yao_2018_b}. The term skin effect refers to the wave function localizing on one boundary (skin) of the lattice. 

Here we observe a non-hermitian skin effect in the $\mathcal{PT}$-broken phase in which, as time evolves, the original-basis state $|\psi(t)\rangle=e^{-iHt}|\psi(t=0)\rangle$ becomes localized around the first site at the left edge of the tight-binding lattice. This localization may also be seen in the Krylov space in the following way. The dynamics of wave function spreading in position basis may be mapped to the dynamics of a dual tight-binding chain in Krylov space, constructed from a given generic initial state $|\psi(t=0)\rangle$ \cite{Balasubramanian:2022tpr}, with the Lanczos coefficients providing the couplings for the tight-binding chain in Krylov space. Localization in the lattice site basis corresponds to localization in the Krylov basis through a change of basis vectors.

In the $\mathcal{PT}$-unbroken phase, the time-evolved single-particle state $|\psi(t)\rangle$ spans the entire system and oscillates across different sites over time in both the original and the dual Krylov lattice. The spread complexity, defined on the dual Krylov basis, is a distinctive indicator of this phase transition. In the $\mathcal{PT}$-unbroken scenario, when all the eigenvalues are real, there is an initial growth of complexity, followed by a saturation phase with large oscillations. These oscillations can be attributed to the unitary dynamics and finite-size effects of the tight-binding chain \cite{Pal:2023yik}. In contrast, in the $\mathcal{PT}$-broken case, spread complexity shows a sharp growth followed by stable saturation. Moreover, compared to the $\mathcal{PT}$-unbroken phase, the saturation value of complexity is significantly suppressed in the broken phase. This suppression of the complexity saturation value is due to the localization of the state along non-unitary time evolution. As the state is localized in the Krylov basis, it has less number of supports on Krylov basis vectors. This results in the suppression of complexity. These results reflect how complexity dynamics of quantum state evolution, phrased using the Krylov space approach captures the $\mathcal{PT}$-phase transition and non-hermitian skin effect. 

To systematically quantify the strength of localization in the dual Krylov basis, we use a dynamic measure, the Krylov inverse participation ratio (KIPR), inspired by the inverse participation ratio (IPR) \cite{Edwards_1972, Misguich_2016} for Hermitian evolution. IPR in Hermitian systems is known as a good measure to probe Anderson localization. Unlike the conventional IPR, KIPR serves as a dynamic quantifier. A higher value of KIPR indicates stronger localization in the Krylov space. The strength of localization varies for different values of the critical parameter and the spread of the initial state on the tight-binding lattice. In \cite{Bento:2023bjn}, KIPR is used to study localization in a Hermitian model. Our work defines KIPR for non-Hermitian Hamiltonian dynamics. This definition is different from KIPR in the Hermitian setup since, for a non-Hermitian Hamiltonian, the eigenstates do not form an orthogonal set of basis vectors. Nonetheless, using a version of the Lanczos algorithm adapted to complex symmetric Hamiltonians \cite{Bhattacharya:2023yec}, we build a complex orthogonal Krylov basis that is applicable to $\mathcal{PT}$-symmetric systems. We note that in \cite{beetar2023complexity} $\mathcal{PT}$-phase transitions were studied using Krylov operator complexity.

\paragraph*{Main result.---} We find a localization-delocalization transition along with the $\mathcal{PT}$-phase transition in the model described by the Hamiltonian Eq.~\eqref{eq:PT_Hamiltonian}. In the $\mathcal{PT}$-broken phase, the time-evolved wave function localizes at the left-most lattice site of the tight-binding chain, manifesting the non-hermitian skin effect.  We find that the saturation values of spread entropy and entropic complexity experience a suppression proportional to the localization strength quantified by the KIPR. In the $\mathcal{PT}$-broken phase, we find that this suppression of the saturation value of spread complexity is not correlated with the localization strength. In this phase, spread complexity is no longer a reliable probe of the strength of localization, unlike in the unitary phase \cite{Rabinovici:2021qqt}. However, in this phase, entropic complexity still reflects the localization strength: it measures the effective Hilbert space dimension required to store the entropy distribution and is, therefore, highly sensitive to the localization strength.

\paragraph*{\texorpdfstring{$\mathcal{PT}$}~ phase transition and localization.---}

\begin{figure}[hbtp]
         \centering
         \includegraphics[width=0.48\textwidth]{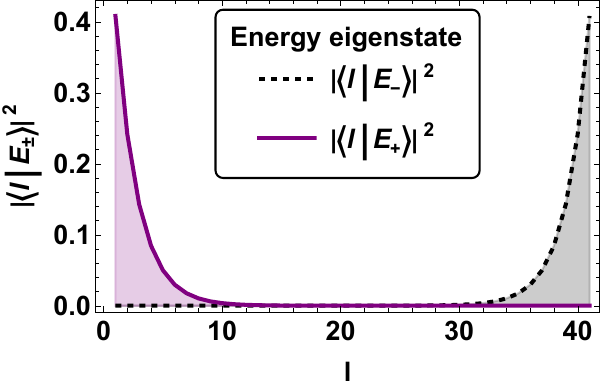}
         \caption{Right (black dashed) and left (purple) edge states in the $\mathcal{PT}$-broken phase with $N=40$, $J=1$ and $\gamma=1.3$, corresponding to energy eigenvalues $E_{\kappa-}=-0.530769 i$ and $E_{\kappa+}=0.530769 i$, respectively. We see that the edge states $|E_{\pm}\rangle$ are localized on the left and right edges of the tight-binding lattice. }\label{fig:edge_state}
\end{figure}

We discuss the phase transition properties of the system governed by the $\mathcal{PT}$-symmetric tight-binding Hamiltonian of Eq.~\eqref{eq:PT_Hamiltonian} and define the spread complexity. This Hamiltonian has uniform nearest-neighbour hopping interactions and features two complex conjugate imaginary on-site potentials, denoted as $\pm i\gamma$. The tunnelling strength $J$ is set to one throughout this paper. The action of the parity operator $\mathcal{P}$ and the time-reversal operator $\mathcal{T}$ is given by $\mathcal{T}i\mathcal{T}=-i$ and $\mathcal{P}a_l^\dagger\mathcal{P}=a_{N+1-l}^\dagger$. The Hamiltonian exhibits $\mathcal{PT}$ symmetry, i.e, $\left[H,\mathcal{PT}\right]=0$. The $\mathcal{PT}$-unbroken phase corresponds to the case when all eigenstates of the Hamiltonian given in Eq.~\eqref{eq:PT_Hamiltonian} are also eigenstates of the $\mathcal{PT}$ operator. In this case, all the eigenvalues are real, and the evolution is unitary. In the $\mathcal{PT}$-broken phase,  $N-2$ eigenvalues are real, and two eigenvalues form a complex-conjugate pair \cite{Jin:2009zc},
\begin{align}
& E_{\kappa \pm} = \pm i 2J \sinh(\kappa).
\end{align}
Here $\kappa$ satisfies the transcendental equations

\begin{align}
    \gamma^2  = \begin{cases}J^2 \frac{\sinh \left(\kappa \left( N+1 \right) \right)}{\sinh \left(\kappa \left( N-1 \right) \right)}~~~ \text{for}~~~ N=2n+1,\\
 J^2 \frac{\cosh \left(\kappa \left( N+1 \right) \right)}{\cosh \left(\kappa \left( N-1 \right) \right)}~~~ \text{for}~~~ N=2n.
 \end{cases}
\end{align}
We find that the eigenvectors corresponding to the complex energy eigenvalues $E_{\kappa+}$ and $E_{\kappa-}$ are the left edge state $|E_+\rangle$ and the right edge state $|E_-\rangle$, respectively. As shown in Fig.~\ref{fig:edge_state}, $|E_+\rangle$ and $|E_-\rangle$ are localized around the left and right edge of the lattice, respectively. These states are not eigenstates of $\mathcal{PT}$ operator, thus breaking the $\mathcal{PT}$-symmetry. The skin effect refers to the time-evolved wave function localizing on one boundary or the skin of the lattice. 

As discussed in \cite{Jin:2009zc, Graefe_2008}, the transition between the $\mathcal{PT}$-symmetric and broken phases occurs at the following parameter values of the imaginary coupling $\gamma$ of the Hamiltonian Eq.~\eqref{eq:PT_Hamiltonian},
\begin{align} \label{eq:transition_pt}
\gamma=
\begin{cases}
    J \sqrt{(n+1)/n} ,~~~~\text{for}~~~~~~ N &=~2n+1, \\
    J,~~~~~~~~~~~~~~~~~~~~\text{for}~~~~~~ N&=~2n.
\end{cases}
\end{align}
As shown in \cite{Jin:2009zc}, at the transition point Eq.~\eqref{eq:transition_pt}, two energy eigenvalues become zero, corresponding to exceptional points \cite{Jin:2009zc, Graefe_2008}.

 We now apply the Krylov state complexity \cite{Balasubramanian:2022tpr} to our model. Starting from an initial state $|\psi(0)\rangle = |K_0\rangle$, we construct the Krylov basis $\{ | K_n\rangle\}$ using an adaptation of Lanczos algorithm to complex symmetric Hamiltonians \cite{200SIAM, Bhattacharya:2023yec}, such as Eq.~\eqref{eq:PT_Hamiltonian}.  The Krylov basis vectors form a matrix $Q$, which follows the complex orthogonality relation, $Q^TQ=I_n$. In this basis, the Hamiltonian takes a tri-diagonal form as $\tilde{T}=Q^THQ$. The complex symmetric Lanczos algorithm is reviewed in Appendix \ref{lanc}. The time-evolved state $|\psi(t)\rangle$ may be expanded in the Krylov basis $|K_n\rangle$ as $|\psi(t)\rangle= \sum_{n} \tilde{\psi}_n(t) |K_n\rangle$ and the spread complexity and entropy are defined by
\begin{equation} 
    C(t)=\sum_{n} np_n,~S(t) = -\sum_n p_n \ln p_n.\label{original_def}
\end{equation}
Here $p_n=|\tilde{\psi}_n(t)|^2$ and we dynamically normalize the time-evolved state by $\mathcal{N}=\sqrt{\text{Tr}[e^{-i H t}|\psi(0)\rangle\langle\psi(0)|e^{i H^{\dagger} t}]}$ to always fix the total probability to one.

An entropic definition of complexity is given by \cite{Balasubramanian:2022tpr}
\begin{equation}
C_S(t) = e^{S(t)}\label{alt_def},
\end{equation}
with $S(t)$ given by Eq.~\eqref{original_def}. This entropic complexity is independent of the specific choice of monotonic weighting used in defining the spread complexity in Eq.~\eqref{original_def}. It quantifies the minimal Hilbert space dimension necessary to accommodate the probability distribution of Krylov basis weights. 

We proceed by discussing the localization-delocalization transition that the system Eq.~\eqref{eq:PT_Hamiltonian} undergoes
along with the $\mathcal{PT}$-phase transition.  To dynamically assess the strength of wave function localization in the dual Krylov basis, we introduce the \textit{Krylov inverse participation ratio} (KIPR),

\begin{equation}\label{kipr}
    \text{KIPR}(t)= \sum_{n=0}^{D-1} |\langle K_n|\psi(t)\rangle|^4 = \sum_{n=0}^{D-1}|\tilde{\psi}_n(t)|^4.
\end{equation}
This measure acts as a dynamic probe of the wave function localization in the Krylov basis over time. The KIPR$(t)$ is one when the state is fully localized as $|\psi(t)\rangle=|K_n\rangle$ for a specific $n$. In cases when the time-evolved state has support on all sites, $|\psi(t)\rangle=\frac{1}{\sqrt{D}}\sum_{j=0}^{D-1}|K_j\rangle$, the KIPR is  given by ${1}/{D}$ in 
$D$-dimensional Krylov space.

\paragraph*{Localization in $\mathcal{PT}$-broken phase.---}
We now show that in the $\mathcal{PT}$-broken phase, the wave functions localize in the lattice basis, giving rise to the non-hermitian skin effect.
The amplitude of the energy eigenstate $|E_+\rangle$ corresponding to the energy eigenvalue $E_{\kappa+}=i2J\sinh{\kappa}$ grows exponentially with time, 
\begin{align}
   |E_+(t)\rangle = e^{-iHt} |E_+(0)\rangle=e^{ 2J \sinh(\kappa) t } |E_+(0)\rangle.   
\end{align}
At late times, the maximum contribution in a time-evolved wave function $\psi(t)$ comes from the state $| E_+ \rangle$. The left edge state $| E_+ \rangle$ has maximum support on the first site of the actual lattice, as shown in Fig.~\ref{fig:edge_state}. Hence, at late times, a generic state localizes at the first site of the lattice. Localization in the tight-binding lattice basis is shown in Fig.~\ref{fig:localization_lattice} in Appendix \ref{localization_basis}. This localization is known in the literature \cite{Yao_2018, Yao_2018_b, skinedge} as the non-hermitian skin effect. On the other hand, in the $\mathcal{PT}$-unbroken phase, where all the eigenvalues are real, the wave function does not localize and shows oscillatory behaviour spanning all the lattice. 

\begin{figure}[hbtp]
         \centering
\includegraphics[width=0.48\textwidth]{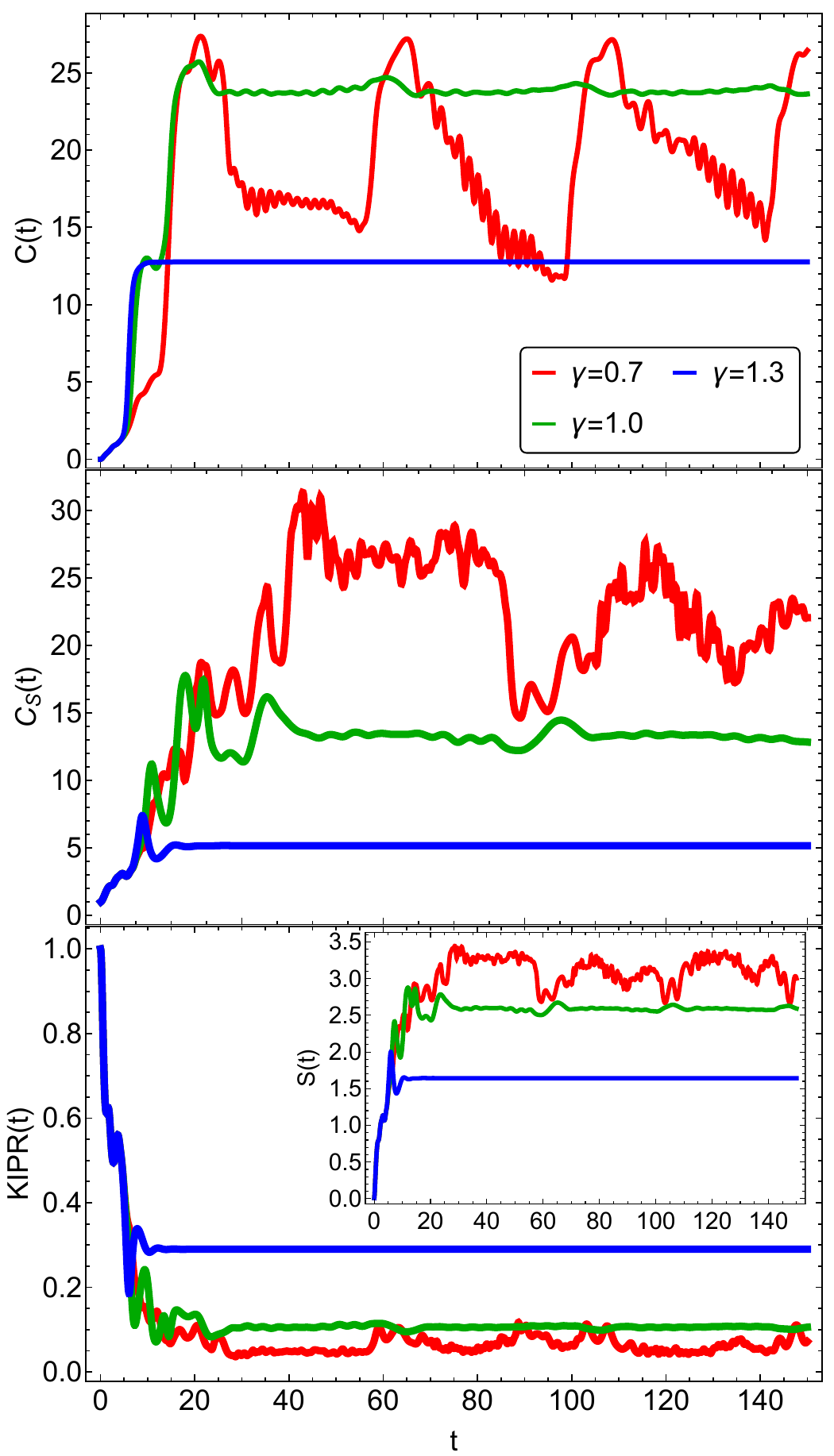}
         \caption{Time evolution of spread complexity (top), entropic complexity (middle), Krylov inverse participation ratio (bottom) and spread entropy (inset) in different phases for the initial state spread over $12^{th}$-$ 18^{th}$ sites of the chain with $N=40$ sites and $J=1$. Spread complexity and entropy saturation values are suppressed in the $\mathcal{PT}$- broken phase compared to the symmetric phase. In the $\mathcal{PT}$-broken phase higher value of localization is measured by KIPR compared to the symmetric phase.}\label{fig:transition}
\end{figure}

This localization-delocalization transition can also be seen in Krylov space. In Fig.~\ref{fig:transition}, we plot the time evolution of spread complexity, entropic complexity and KIPR  defined in Eq.~\eqref{original_def}, Eq.~\eqref{alt_def} and Eq.~\eqref{kipr} in different phases for the initial state spread over $12^{th}$-$ 18^{th}$ sites of the chain with $N=40$ sites. Different phases can clearly be distinguished. In the $\mathcal{PT}$-symmetric phase $(\gamma <1 )$, the spread complexity shows an initial growth followed by large oscillations that are caused by the real spectra and by finite-size effects. In contrast to the symmetric phase, we find that in the $\mathcal{PT}$-broken phase, $\gamma >1$, the spread complexity saturates rapidly following a smaller but faster initial rise than in the $\mathcal{PT}$-symmetric case.  At $\gamma = 1$, there is a critical point at which the spread complexity shows lower initial growth and higher saturation value compared to the symmetric and broken phases, respectively. Moreover, the fluctuations are smaller at the critical point than in the $\mathcal{PT}$-symmetric phase due to the transition to the localized regime. 

The spread complexity of a state measures how much the state spreads in the Krylov basis under the time evolution. For finite-size systems, since the number of steps in the Lanczos algorithm is bounded by the dimension of the Hilbert space, the support of the time-evolved state in the Krylov basis stops growing once the state has explored the full Krylov space. After the time required to explore the full Krylov space, the spread complexity either saturates or oscillates. The saturation of complexity implies that the support of the time-evolved state has reached a steady number of Krylov basis vectors. In the $\mathcal{PT}$-symmetric phase, the time-evolved state is delocalized in the Krylov space and has larger support in the Krylov basis as compared to the $\mathcal{PT}$-broken phase. In the broken phase, the state localizes in the Krylov space and, therefore, has a smaller complexity saturation value.

 Now we turn to entropic complexity defined in Eq.~\eqref{alt_def}. This may also be used to probe the $\mathcal{PT}$- phase transition. During the phase transition, its qualitative behaviour is similar to the spread complexity. The time dependence of the entropic complexity is shown in the central panel of Fig.~\ref{fig:transition} for the same initial state as for the spread complexity. The entropic complexity also shows a lower value in the broken phase as compared to the symmetric case, which again indicates localization of the time-evolved state in the $\mathcal{PT}$-broken phase. As we will discuss below, in the broken phase, the time evolution is different for spread and entropic complexities.

The KIPR is plotted in the lowest panel of Fig.~\ref{fig:transition} for the initial state spread around the centre of the tight-binding chain. Since at $t =0$, the initial state is fully localized at a specific vector as $| \psi(0) \rangle = | K_0 \rangle$, the KIPR is one at $t=0$. In the symmetric phase, the KIPR decays and eventually oscillates around zero, implying that localization is absent. On the other hand, in the broken phase, the KIPR shows decay followed by a ramp and finally saturates at a constant value. A higher saturation value in KIPR indicates stronger localization, as explained in the paragraph following Eq.~\eqref{kipr}.  In the inset of Fig.\ref{fig:transition}, we plot the spread entropy for the same initial state as for the other three measures. The late-time value of the KIPR at the critical point lies between the ones for the $\mathcal{PT}$-unbroken and broken phases. Since the KIPR measures localization, the higher saturation value of the KIPR in the broken phase compared to the symmetric phase implies that the $\mathcal{PT}$-symmetry transition and the localization-delocalization transition (LDT) occur in parallel in the system considered, giving rise to a non-hermitian skin effect.

\begin{figure}[hbtp]
         \centering
         \includegraphics[width=0.48\textwidth]{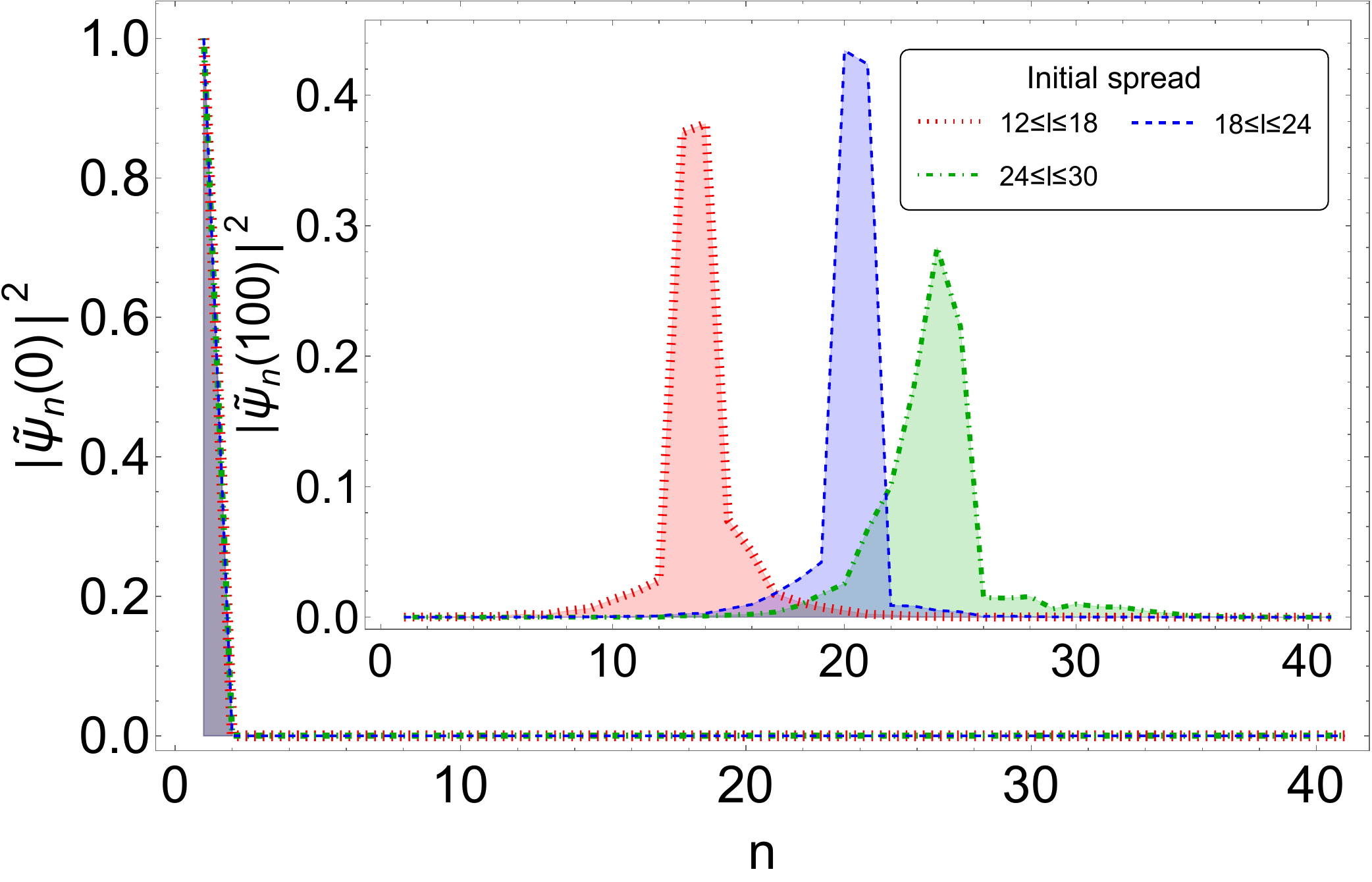}
         \caption{Localization of time-evolved initial states with different spreads on the Krylov basis at time $ t=0$ (main plot) and $t=100$ (inset plot). We choose the total number of sites $N=40$, $J=1$, and $\gamma = 1.3$. The localization of the wave function in the Krylov basis is determined by the spread of the initial state in the tight-binding chain.}\label{fig:localization_K_space}
\end{figure}

The place of localization within the Krylov space can be precisely determined based on the choice of the initial state’s spread. We plot the amplitude of the Krylov wave function, $|\tilde{\psi}_n(t)|^2$, for different initial states at initial and late times (inset) in Fig.~\ref{fig:localization_K_space}. At time $t=0$, as expected, we notice all the initial states completely localized in the first Krylov vector as $| \psi(0)\rangle = | K_0\rangle$.  However, when the initial state is spread between sites $l_1$ and $l_2$ of the lattice chain, the resulting time-evolved state becomes localized on the Krylov basis vectors $| K_n \rangle$ with $l_1 \leq n \leq l_2$ at late times. This phenomenon is represented in the inset of Fig.~\ref{fig:localization_K_space}. The amplitudes of the overlap of a time-evolved state on the Krylov basis at late times are not unique for all initial states. Instead, they depend on the position of the initial states. When the spread of the initial state includes the central site, the amplitude of the overlap is higher compared to when this is not the case, as shown in the inset of Fig.~\ref{fig:localization_K_space}. See Appendix \ref{localization_saturation} for further details about the dependence of the saturation value of complexity on different initial spreads.

\begin{figure}[hbtp]
         \centering         \includegraphics[width=0.48\textwidth]{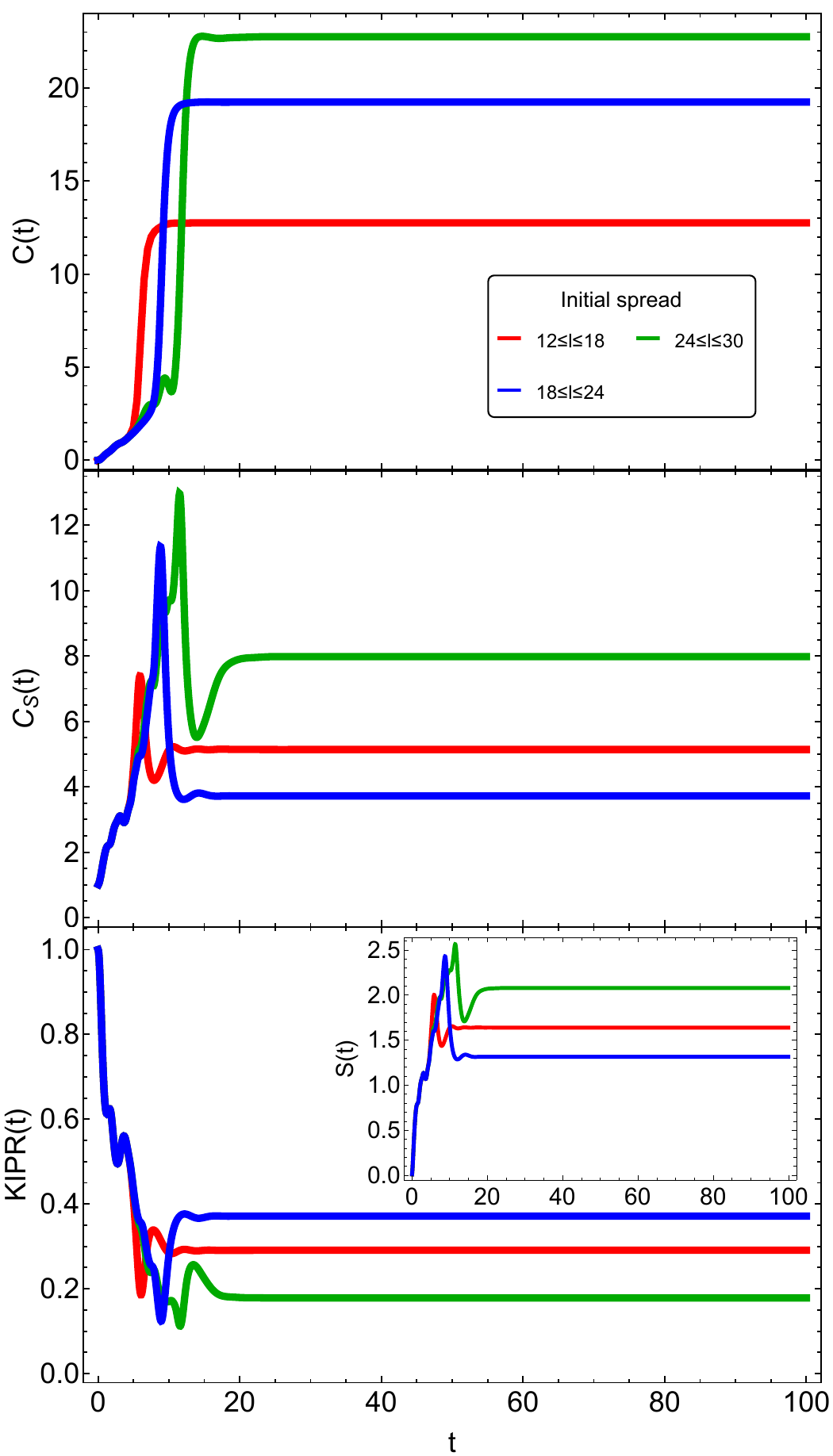}
         \caption{Saturation of spread complexity (top), entropic complexity (middle), KIPR (bottom) and spread entropy (inset) for different initial states in broken phase. We choose the total number of sites $N=40$, tunnelling strength $J=1$, and $\gamma = 1.3$. The hierarchy of saturation values of complexity does not necessarily follow from the strength of localization; instead, the saturation value of spread entropy is suppressed accordingly.}\label{fig:saturation}
\end{figure}

According to our findings shown in Fig.~\ref{fig:saturation}, the value at which spread complexity, entropic complexity, and KIPR saturate significantly depends upon the precise location of the initial state. For the initial state spread over the central site of the chain as shown in Fig.~\ref{fig:localization_K_space}, the corresponding saturation value of the spread complexity as shown by the blue curve in the first panel of Fig.~\ref{fig:saturation} occupies an intermediate position, even if the localization strength is the highest. In contrast, the suppression of entropic complexity and spread entropy for different initial states always occurs in accordance with the strength of localization, as can be seen from the middle and bottom panels of Fig.~\ref{fig:saturation}. This observation supports the finding that for the $\mathcal{PT}$-broken phase, stronger localization does not always imply stronger suppression of spread complexity. However, spread entropy and entropic complexity are always suppressed inversely proportional to the strength of localization.
The entropic complexity given in Eq.~\eqref{alt_def} measures the dimension of the minimum Hilbert space required to contain the information of the time-evolved state. In the $\mathcal{PT}$-broken phase, the effective Hilbert space dimension is reduced proportionally to the localization strength. Therefore, we conclude that in the presence of localization, the entropic complexity is a better probe to measure the spread of the state as compared to the spread complexity in Eq.~\eqref{original_def}.

This different behaviour of spread and entropic complexity can be understood as follows. Due to the skin effect in the $\mathcal{PT}$-broken phase, the wave function in the position basis is found to localize always at the first site in lattice site basis. However, depending on the choice of initial state, this first position basis corresponds to different Krylov basis vectors $|K_m\rangle$. The further the initial state is from the first lattice site, the higher the site number $m$ where the wave function localizes in Krylov space as shown in Fig.~\ref{fig:localization_K_space}. So, for different initial states corresponding to the same strength of localization, the spread complexity will be higher for the state localized furthest from the initial Krylov vector. This is due to the fact that in the definition of the spread complexity in Eq.~\eqref{original_def}, the Krylov basis vectors with higher site numbers contribute with higher weight factor $n$. The dependence on the position of localization in the Krylov basis causes the spread complexity to lose its sensitivity to the strength of localization in the $\mathcal{PT}$-broken case. Therefore, we infer that for quantum state evolution in the Krylov basis, increasing localization does not necessarily imply suppression of the spread complexity in the non-unitary $\mathcal{PT}$-broken phase. However, spread entropy and the entropic complexity as defined in Eq.~\eqref{alt_def}, are always suppressed for stronger localization.

\textit{Conclusion.---} For a tight-binding Hamiltonian with complex on-site potentials at the edge sites, we have shown that in the $\mathcal{PT}$-symmetric phase, the wave function is delocalized and has oscillatory behaviour at late times.  On the other hand, in the $\mathcal{PT}$-broken phase, the wave function is localized at the left edge site of the tight-binding lattice for the edge state corresponding to the positive-imaginary eigenvalue. This is the main feature of the non-Hermitian skin effect. The skin effect occurs due to the complex energy eigenvalues. Their corresponding eigenstates are localized at the edge of the tight-binding lattice space. The skin effect entails a localization-delocalization transition that occurs simultaneously with the $\mathcal{PT}$-phase transition. In this work, we studied the spread complexity and spread entropy as a means of characterizing this dynamical behaviour. All four Krylov space probes we use in this work, namely, spread complexity, spread entropy, entropic complexity, and KIPR, detect the $\mathcal{PT}$-phase transition. Due to localization, the saturation values of spread complexity, spread entropy and entropic complexity are suppressed in the $\mathcal{PT}$-broken phase. This result for the non-unitary evolution and PT-symmetric systems adds to the general insight \cite{Rabinovici:2021qqt} that localized quantum states are less
complex. 

 The Krylov inverse participation ratio (KIPR) defined in Eq.~\eqref{kipr} acts as a quantifier for the localization strength in Krylov space. In the $\mathcal{PT}$-symmetric phase, when there is no localization, the KIPR value oscillates around $1/D$ where $D$ is the dimension of the Krylov space. In the $\mathcal{PT}$-broken phase, the localization strength depends upon the strength of the on-site imaginary potential, as well as on the spread of the initial state on the tight-binding chain. The spread entropy and entropic complexity are suppressed monotonically with the localization strength. On the other hand, the spread complexity is not correlated with the localization strength. This insensitivity is due to the position dependence of the spread complexity. 

\begin{acknowledgments}
\vspace{3em}
\noindent The authors would like to thank Vijay Balasubramanian, Souvik Banerjee, Pablo Basteiro, Lorenzo Crippa, Diptarka Das, Saskia Demulder, Giuseppe Di Giulio, Mario Flory, Domenico Giuliano, René Meyer, Sara Murciano, Kunal Pal, Kuntal Pal, Subir Sachdev, Giorgio Sangiovanni and Aninda Sinha for useful discussions and comments. The work of A.B.~is supported by the Polish National Science Centre (NCN) grant 2021/42/E/ST2/00234. R.N.D.~and J.E.~are supported by Germany's Excellence Strategy through the W\"urzburg‐Dresden Cluster of Excellence on Complexity and Topology in Quantum Matter ‐ ct.qmat (EXC 2147, project‐id 390858490),
and  by the Deutsche Forschungsgemeinschaft (DFG) 
through the Collaborative Research centre ``ToCoTronics", Project-ID 258499086—SFB 1170. R.N.D.~further acknowledges the support by the Deutscher Akademischer Austauschdienst (DAAD, German Academic Exchange Service) through the funding programme, ``Research Grants - Doctoral Programmes in Germany, 2021/22 (57552340)". This research was also supported in part by the Perimeter Institute for Theoretical Physics. Research at Perimeter Institute is supported by the Government of Canada through the Department of Innovation, Science and Economic Development and by the Province of Ontario through the Ministry of Research, Innovation and Science. B.D.~acknowledges MHRD, India for Research Fellowship. B.D.~would also like to acknowledge the support provided by Max Planck Partner Group grant MAXPLA/PHY/2018577. B.D. would further like to acknowledge the support provided by the MATRICS grant SERB/PHY/2020334.
\end{acknowledgments}

\newpage

\onecolumngrid
\renewcommand{\theequation}{S.\arabic{equation}}
\setcounter{equation}{0}

\section*{Supplemental Material}
\section{Lanczos algorithm adapted to complex symmetric operator}\label{lanc}

We review the algorithm for constructing the Krylov basis associated with a complex symmetric operator $A$ \cite{200SIAM, Bhattacharya:2023yec}. The complex symmetry refers to the property, $A^{T} = A$ but $A^{\dag} \neq A$. The utilization of complex symmetry leads to a reduction in both computational workload and storage requirements when compared to the bi-Lanczos method. Notably, the algorithm demonstrates that for a non-hermitian but complex symmetric operator, only one set of Krylov basis vectors is sufficient for tridiagonalizing the operator, in contrast to the bi-Lanczos algorithm, which demands two sets of basis vectors.
The diagonalizability of a complex symmetric matrix A depends on the feasibility of selecting its eigenvector matrix Z such that $Z^TAZ=\text{diag}(\lambda_1,\lambda_2,...,\lambda_n)$.
The matrix of eigenvectors $Z$, adheres to the condition $Z^TZ=I_n$, indicating complex orthogonality. The importance of complex orthogonality plays a pivotal role in facilitating an efficient construction of the Krylov basis. To ensure this complex orthogonality, even in the presence of complex vectors, a modified Lanczos algorithm is employed.

The introduced \textit{complex symmetric Lanczos} algorithm, as detailed in \cite{200SIAM}, is adept at deriving the tri-diagonal form of a complex symmetric matrix A \cite{Bhattacharya:2023yec}. This algorithm constructs an orthogonal basis ${ | q_j \rangle }$ spanning the Krylov space $\mathcal{K}^j (A, |q_1 \rangle) \equiv {A | q_1 \rangle, A^2 | q_1 \rangle, \dots }$. The basis is initiated with a normalized vector, chosen as the initial state $|q _1 \rangle=|\psi(0)\rangle$. Complex orthogonality of the basis vectors is ensured, with $\langle q_j |q_j \rangle = \delta_{i,j}$, where $\langle q_j | = (|q _j \rangle )^T$. The basis construction involves a three-term recursion relation 
\begin{align}
    \beta_{j+1} |q_{j+1} \rangle = A |q_{j} \rangle - \alpha_j |q_{j} \rangle - \beta_{j} |q_{j-1} \rangle, \label{recursion3}
\end{align}
where $\beta_n=\langle q_{n-1}|A|q_n\rangle$ and $\alpha_n =\langle q_n| A | q_n \rangle $. In the resulting basis ${ | q_j \rangle }$, the complex symmetric matrix A adopts the tri-diagonal form denoted as $\Tilde{T}_j$,
\begin{equation}
\Tilde{T}_j = \begin{pmatrix}
    \alpha_1 & \beta_2 & 0        & \dots     & 0   \\
     \beta_2 & \alpha_2           & \beta_3   &    &  \vdots  \\
     0 &  \ddots     &    \ddots      & \ddots    &     \\
\vdots &       & \beta_{j-1}  & \alpha_{j-1}    & \beta_j   \\
     0 & \dots  &     0   &     \beta_j   & \alpha_j
  \end{pmatrix}.\label{eq:ttilde_matrix}
\end{equation}

Through the utilization of the Lanczos algorithm, which enforces complex orthogonality, the Krylov basis matrix set is obtained as $Q=[|q_1\rangle~|q_2\rangle~...~|q_n\rangle]$, satisfying $Q^TQ=I_n$. This process yields the tri-diagonal form of matrix A in this basis, expressed as $\Tilde{T}_j=Q_j^TAQ_j$.

\section{Relation between localization and saturation of complexity and entropy in the \texorpdfstring{$\mathcal{PT}$}~-broken phase}\label{localization_saturation}
\begin{figure}
         \centering
         \includegraphics[width=\textwidth]{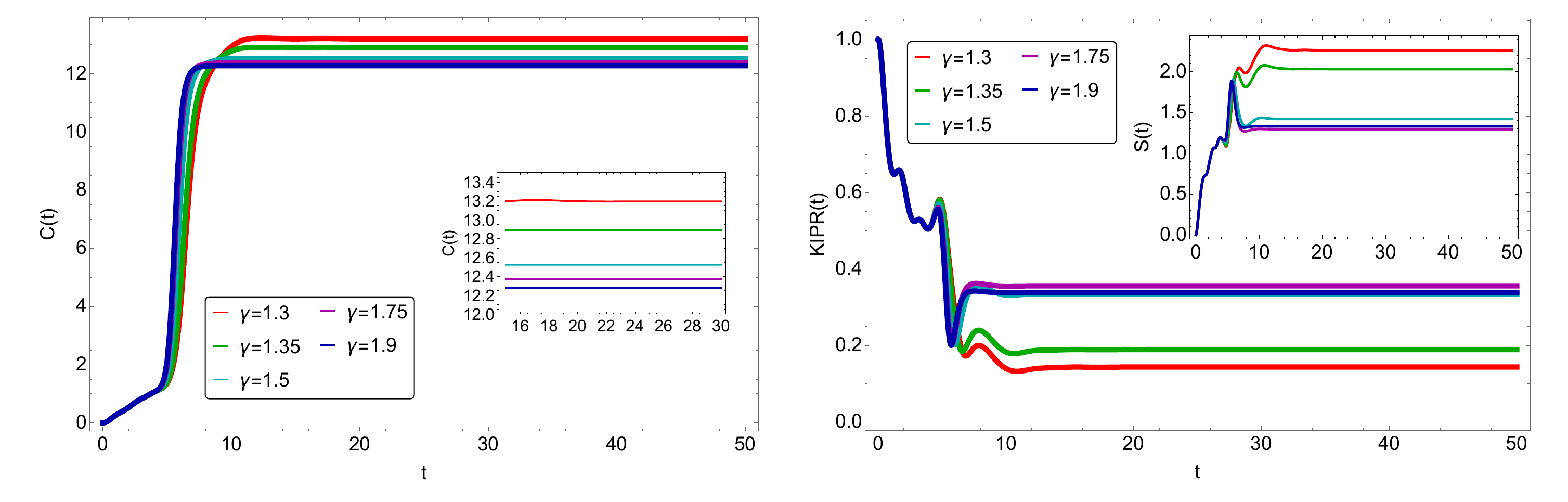}
         \caption{Time dependence of the spread complexity (left), KIPR (right) and spread entropy (inset)  in broken phase for different $\gamma$. The chosen initial state is spread over $12^{th}$ - $19^{th}$ sites of the lattice chain consists of $N = 40$ sites. }\label{fig:broken}
\end{figure}
In Fig.~\ref{fig:broken}, we plot spread complexity, spread entropy, and KIPR for different $\gamma$ in the $\mathcal{PT}$-broken phase $(\gamma>1)$ with the same initial state spread over $12^{th}$ - $19^{th}$ sites of the lattice chain, which consists of $N = 40$ sites. The goal is to understand if stronger localization of the time-evolved state in the Krylov space always necessarily means suppression of complexity and entropy saturation values. We notice that the saturation value of the spread complexity decreases as $\gamma$ increases. However, for spread entropy, the saturation value falls when $\gamma$ rises up-to $\gamma =1.75$. After $\gamma > 1.75$, the saturation value of entropy begins to rise with $\gamma$. The figure of KIPR provides a valuable comparison between the saturation value of spread complexity and entropy with the strength of localization. We observe that there is also a crossover for the saturation value of the KIPR at $\gamma =1.75$, which is perfectly consistent with the spread entropy case. Therefore, spread entropy in the Krylov basis for a quantum state is always more suppressed when localization is stronger; this is not necessarily the case for spread complexity.

\begin{figure}[H]
    \centering
    \includegraphics[width=0.8 \textwidth]{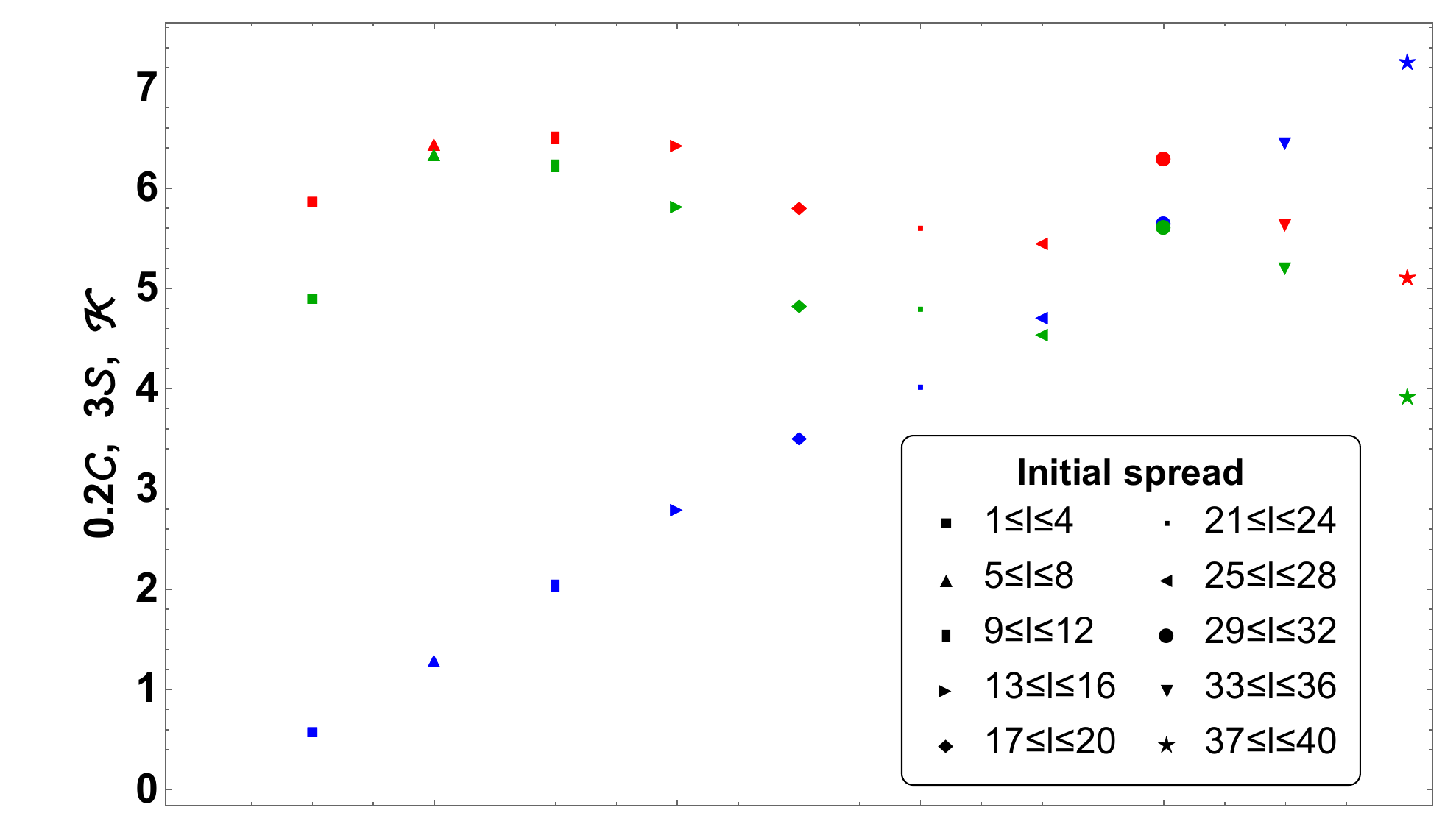}
    \caption{Comparison between the late-time average saturation values of spread entropy (red), the inverse of KIPR (green), and spread complexity (blue) for different choices for the spread of the initial state. Notice that the spread entropy and KIPR follow the exact same pattern while the spread complexity saturation value keeps growing as the initial spread is taken far from the first position basis.}
    \label{fig:satvals}
\end{figure}

The saturation value of spread complexity, entropy, and KIPR depends on the choice of the initial state, as depicted in Fig.~\ref{fig:saturation}. We further validate this pattern by varying the initial state spread over four sites from the left to the right of the chain and observing the dependence of the saturation value of complexity and entropy on changes in the saturation value of KIPR. The saturation value of KIPR and the saturation value of spread entropy should exhibit an inverse relationship, demonstrating that a stronger saturation indicates the suppression of spread entropy. Thus, higher saturation values of KIPR lead to lower saturation values of spread entropy. For visualization, we define $\mathcal{C}$ and $\mathcal{S}$ as the average saturation values of spread complexity and spread entropy, respectively, as
\begin{equation}
    \mathcal{C}= \lim_{t^*\to \infty} \frac{1}{t^*-t_{\text{sat}}} \int_{t_{\text{sat}}}^{t^*} C(t)dt,~~\mathcal{S}= \lim_{t^*\to \infty} \frac{1}{t^*-t_{\text{sat}}} \int_{t_{\text{sat}}}^{t^*} S(t)dt.
\end{equation}

We additionally define $\mathcal{K}$ as the average of the inverse of the KIPR saturation value as 
\begin{align}
\mathcal{K}= \lim_{t^*\to \infty} \frac{1}{t^*-t_{\text{sat}}} \int_{t_{\text{sat}}}^{t^*} \frac{1}{\text{KIPR}(t)}dt.
\end{align}
Here, $t_{\text{sat}}$ is the time when complexity, entropy and KIPR values get saturated and for a particular initial state, this time is same for all these three quantities. Following our modified proposal, $\mathcal{S}$ should be proportional to $\mathcal{K}$. This behaviour aligns precisely with our findings in Fig.~\ref{fig:satvals}. The blue, red and green dots represent the $\mathcal{C}$, $\mathcal{S}$ and $\mathcal{K}$ respectively. We have rescaled $\mathcal{C}$ and $\mathcal{S}$ by 0.2 and 3 times, respectively, for bringing all the values in the same plot. We find $\mathcal{S}$ is suppressed according to the strength of localization as measured by $\mathcal{K}$ but $\mathcal{C}$ keeps growing linearly due to the monotonically increasing weight factor.

\section{Support of the time-evolved states in the tight-binding lattice basis}\label{localization_basis}

\begin{figure}[H]
         \centering   \includegraphics[width=\textwidth]{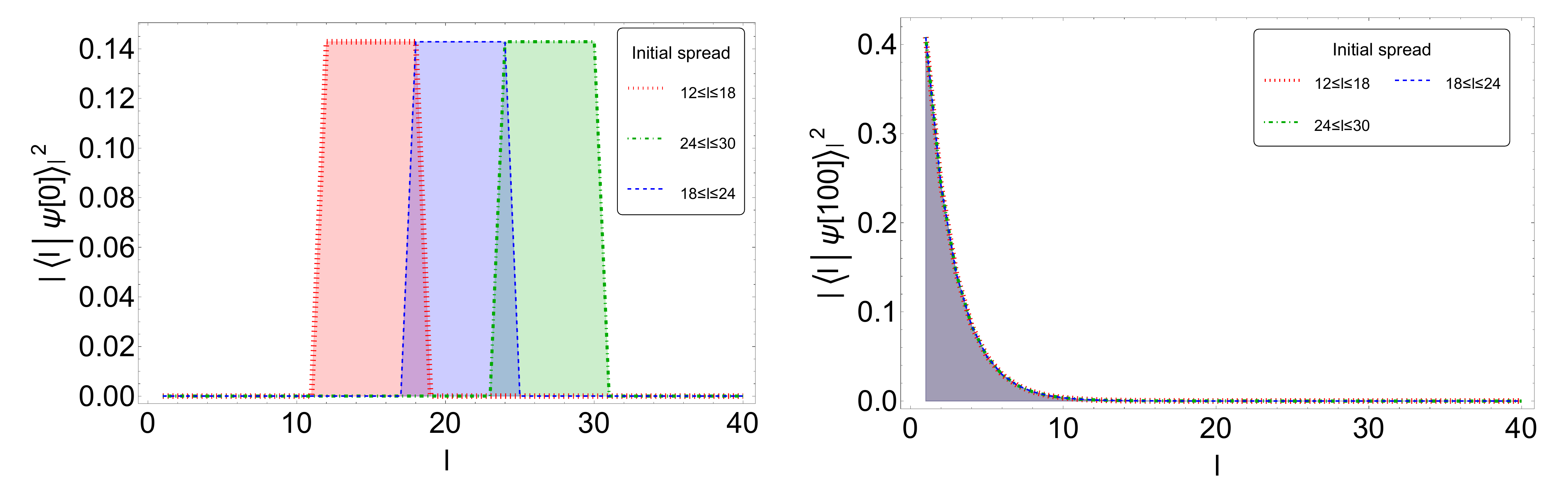}
         \caption{Localization of time-evolved different initial states on lattice sites at initial (left) and late times (right).}\label{fig:localization_lattice}
\end{figure}

In Fig.~\ref{fig:localization_lattice}, we illustrate the localization of the initial states on the lattice basis in the $\mathcal{PT}$-broken phase. We consider three different spreads of the initial state. Initially, the states have equal support on the different lattice sites for $N=40$. The tri-diagonal Hamiltonian in the $\mathcal{PT}$-broken phase has an unbalanced gain parameter in the first diagonal. Due to this unbalanced gain, the state localizes at the first site of the lattice at late times. Correspondingly, the localization of the time-evolved states in the Krylov basis is shown in Fig.~\ref{fig:localization_K_space}. Since all the initial states are taken as the $|K_0\rangle = \delta_{n,0}$, the states are localized in a specific vector in the Krylov basis. The Hamiltonian for an arbitrary initial state changes after performing the Lanczos algorithm, and the new tri-diagonal matrix $\tilde{T}$ that we get in this way would always have the first non-hermiticity present around the sites where the chosen initial state had the first nonzero support. As soon as the Krylov wave function hits these sites, it localizes and stays so afterwards. Therefore, in the $\mathcal{PT}$-broken phase, complexity analysis is dependent on the initial state, and the localization of the time-evolved state on the Krylov basis, always helps us to understand which site exactly the chosen initial state had the first nonzero support on. Comparing Fig.~\ref{fig:localization_K_space} and Fig.~\ref{fig:localization_lattice}, we can see a dual mapping between the localization of the states in the two different bases.  However, the dual mapping is approximate since the amplitudes of the localization of initial states on the lattice at early times are the same. In contrast, the amplitudes of the localization of initial states in the Krylov basis at late times are not the same, see Fig.~\ref{fig:localization_K_space}. These different amplitudes of localization in the Krylov basis are the main reason for the different strengths of localizations.   

\bibliographystyle{apsrev4-1}
\bibliography{apssamp}
\end{document}